\documentclass{article}

\usepackage{arxiv}

\usepackage[utf8]{inputenc} % allow utf-8 input
\usepackage[T1]{fontenc}    % use 8-bit T1 fonts
\usepackage{hyperref}       % hyperlinks
\usepackage{url}            % simple URL typesetting
\usepackage{booktabs}       % professional-quality tables
\usepackage{amsfonts}       % blackboard math symbols
\usepackage{nicefrac}       % compact symbols for 1/2, etc.
\usepackage{microtype}      % microtypography
\usepackage{amsmath}
\usepackage{float}
\usepackage{varioref}

\usepackage{graphicx}
\newcommand{\Mathematica}{\textit{Mathematica\textsuperscript{\resizebox{!}{0.8ex}{\textregistered}}}}

\title{Exact Area Law for Planar Loops in Turbulence in Two and Three Dimensions}

\author{
  Alexander Migdal \\
  Fresnel Research LLC
  } 
\begin{document}
\maketitle

\begin{abstract}
We study properties of the minimal surface in the Area Law Solution \cite{M93}, \cite{M19a}, \cite{M19b}. We find out that Area Law holds exactly for 2D turbulence as well as for arbitrary planar loop in higher dimensions. This relies on our previous result $\alpha = \frac{1}{2}$ in which case the second moment of circulation can be proven to reduce to the area inside the planar loop.
In $d=3$, we demonstrate how the Stokes condition $\partial_i \omega_i(r)=0$ is exactly satisfied for the minimal surface solution in virtue of vanishing mean curvature at the minimal surface. 
In order to satisfy Loop Equation beyond planar loops, we introduce self-consistent conformal metric on the surface designed to preserve Stokes condition but to  compensate the terms in the loop equation.
We derive nonlinear integral equation for this conformal metric as a function of a point on a surface.
\end{abstract}

% keywords can be removed

\section{Introduction}
In the previous two papers \cite{M19a},\cite{M19b} we reviewed and advanced the Minimal Area Solution \cite{M93} for the Loop Equation in turbulence, which was recently verified experimentally \cite{S19}.
Let us repeat the latest revision of this theory before we  start advancing it further.

The basic variable in the Loop Equations a circulation around closed loop in coordinate space

\begin{equation}\label{Gamma}
    \Gamma = \oint_C \vec v d\vec r 
\end{equation}

The PDF  for  velocity circulation as a functional of the loop 
\begin{equation}
    P\left ( C,\Gamma\right) =\left < \delta\left(\Gamma - \oint_C \vec v d\vec r\right)\right>
\end{equation}

with brackets \begin{math}< > \end{math} corresponding to time average or average over random forces, was shown to satisfy certain functional equation (loop equation). 

\begin{equation}\label{LoopEq0}
\frac{\partial}{\partial \Gamma} \frac{\partial}{\partial t}  P\left ( C,\Gamma\right) 
=\oint_C d r_i \int d^3\rho\frac{ \rho_j }{4\pi|\vec \rho|^3}\frac{\delta^2 P(C,\Gamma)}{\delta \sigma_k(r) \delta \sigma_l(r + \rho)}\left(\delta_{i j}\delta_{k l} - \delta_{j k}\delta_{i l}\right)
\end{equation}

The  area derivative is defined using the difference between $P(C+\delta C,\Gamma)- P(C,\Gamma)$ where an infinitesimal loop $\delta C$ around the 3d point $r$ is added as an extra connected component of $C$. In other words, let us assume that the loop $C$ consists of an arbitrary number of connected components $C = \sum C_k$. 
We just add one more infinitesimal loop at some point away from all $C_k$. 
In virtue of the Stokes theorem, the difference comes from the circulation $\oint_{\delta C} \vec v d\vec r$ which reduces to vorticity at $r$
\begin{equation}
  P(C+\delta C,\Gamma)-P(C,\Gamma) =  d\sigma_i(r)    \left <\omega_i(r) \delta'\left(\Gamma - \oint_C \vec v d\vec r\right)\right>
\end{equation}
where
\begin{equation}
    d\sigma_k(r) = \frac{1}{2}\oint_{\delta C} e_{i j k} r_i d r_j
\end{equation}
is an infinitesimal vector area element inside $\delta C$.
In general, for the Stokes type functional, by definition:
\begin{equation}
    U[C+\delta C]-U[C] = d\sigma_i(r) \frac{\delta U[C]}{\delta \sigma_i(r)}
\end{equation}
The Stokes condition $\partial_i \omega_i(r) =0$ translates into
\begin{equation}\label{Stokes}
   \partial_i \frac{\delta U[C]}{\delta \sigma_i(r)} =0
\end{equation}
where $\partial_i = \frac{\partial}{\partial r_i}$ is an ordinary spatial derivative, rather than a singular loop derivative introduced in the Non-Abelian Loop Equations.

The fixed point of the chain of the loop equations (\ref{LoopEq0}) was shown to have a solution which is an arbitrary function of minimal area \begin{math} A_C \end{math} bounded by $C$.
\begin{equation}
    P(C,\Gamma) = F\left(A_C,\Gamma\right)
\end{equation}
The Minimal Area can be reduced to the Stokes functional by the following regularization
\begin{equation}\label{Regularized}
  A_C =\min_{S_C} \int_{S_C} d \sigma_{i}(r_1) \int_{S_C} d \sigma_{j}(r_2) \delta_{i j} \Delta(r_1-r_2)
\end{equation}
with 
\begin{equation}
\Delta(r) = \frac{1}{r_0^2} \exp\left(-\pi \frac{r^2}{r_0^2}\right); r_0 \rightarrow 0
\end{equation}
representing two dimensional delta function, and integration goes over minimized surface $S_C$ (see Figures in \cite{M19b}).

In real world this $r_0$ would be the viscous scale $\left(\frac{\nu^3}{\mathcal {E}}\right)^{\nicefrac{1}{4}}$.
This is a positive definite functional of the surface as one can easily verify using spectral representation:
\begin{equation}
    \int_{S_C} d \sigma_{i}(r_1) \int_{S_C} d \sigma_{j}(r_2) \delta_{i j} \Delta(r_1-r_2) \propto \int d^3k \exp\left(-\frac{k^2r_0^2}{4\pi}\right) \left| \int_{S_C} d \sigma_{i}(r)e^{i k r} \right|^2
\end{equation}
In the limit $r_0 \rightarrow 0$ this definition reduces to the ordinary area:
\begin{equation}
  A_C \rightarrow \min_{S_C} \int_{S_C} d^2 \xi \sqrt{g} 
\end{equation}

\section{Stokes Condition and Mean Curvature}
Let us study deeper this amazing duality of Minimal Surface to the turbulent flow. First of all, why minimal surface -- there is no string theory which would require this minimal surface arise as a classical solution (or at least we do not know any well defined string theory equivalent to Turbulence in spite of some interesting observations \cite{TSVS}).

The Stokes condition (\ref{Stokes}) is satisfied in virtue of minimum condition. When the surface changes into $S'$, the linear variation reduces by volume Stokes theorem to 
\begin{equation}
    \oint_{S'-S} d \sigma_i(r) \frac{\delta U[C]}{\delta \sigma_i(r)} = \int_{\delta V} d^3 r \partial_i \frac{\delta U[C]}{\delta \sigma_i(r)}
\end{equation}
with $\delta V$ being infinitesimal volume between $S'$ and $S$. This linear variation must vanish by definition of the minimal surface,  for regularized area as well as for its local limit.

The area derivative of the Minimal Area in regularized form, then, as before, reduces to elimination of one integration
\begin{equation}\label{vorticity}
  \frac{\delta A_C}{\delta \sigma_i(r)} = 2 \int_{S_C} d \sigma_{i}(\rho) \Delta(r-\rho) \rightarrow 2 n_i(\Tilde{r})\exp\left(-\pi \frac{r_{\perp}^2}{r_0^2}\right)
\end{equation}
Where $n_i(\Tilde{r}) $ is the local normal vector to the minimal surface at the nearest surface point $\Tilde{r}$ to the 3d point $r$, and $r_{\perp}$ is the component normal to the surface at $\Tilde{r}$. With this regularization area derivative is defined everywhere in space but it exponentially decreases away from the surface. Exactly at the surface it reduces to twice the unit normal vector.

Let us investigate this issue in more detail.

Let us choose $x,y$ coordinates in a local tangent plane to the minimal surface at some point which we set as an origin.  In the quadratic approximation (which will be enough for our purpose), equation of the surface reads:
\begin{align}
     z &= \frac{1}{2} \left(K_1 x^2  + K_2 y^2\right)\\
     n &= \frac{\left[ -K_1 x, -K_2 y, 1\right]}{\sqrt{1 +  K_1^2 x^2 + K_2^2 y^2}}\label{Normal2}
\end{align}
where $K_1, K_2$ are main curvatures in planes $y=0, x=0$.

Now, at $r_0 \rightarrow 0$ the Stokes condition reduces to :
\begin{align}
     &\int_{-\infty}^{\infty}d x \int_{-\infty}^{\infty} d y \partial_z \Delta (\vec r)\\
     &\propto
     \int_{-\infty}^{\infty}d x \int_{-\infty}^{\infty} d y \frac{z}{r_0^4} \exp\left(-\pi \frac{x^2+ y^2}{r_0^2}\right)\\ &\propto
     \int_{-\infty}^{\infty}d x \int_{-\infty}^{\infty} d y \frac{K_1 x^2 + K_2 y^2}{r_0^4} \exp\left(-\pi \frac{x^2+ y^2}{r_0^2}\right) \\
     &\propto K_1 + K_2 =0\label{MeanCurvature}
\end{align}
 
This is the mean curvature. So, the Stokes condition is \textbf{equivalent} to the equation for the minimal surface. This is nice to know!
\section{Loop Equation Beyond Logarithmic Approximation}

An important unanswered question in our recent papers \cite{M19a},\cite{M19b} is whether the loop equation was satisfied beyond leading logarithmic approximation.

Within the Area law Anzatz the stationary solution of the loop equation (\ref{LoopEq0}) reduces in the limit $r_0 \rightarrow 0$ to:
\begin{equation}\label{LE1}
    r_0 \oint_C d r_i \int_{S_C} d \sigma(r') \frac{ (r'_i - r_i)n_k(r)n_k(r') - n_i(r')(r'_k - r_k) n_k(r) }{|\vec r' - \vec r |^3}=0
\end{equation} 
Here $d\sigma(r') = \left| d \vec \sigma(r')\right|$ is the scalar area element for the point $\vec r'$ at the surface.  The distance $r_i - r'_i$ is measured in Euclidean space rather than along the surface.
The factor of $r_0$ implies that time derivative of our PDF is very small in the limit when viscosity goes to zero. In general $d-$ dimensional problem it would be $r_0^{d-2}$ and in particular, in two dimensions there would be no factor at all.
By renormalizing time:
\begin{equation}
    t = \tau T_0; \, T_0 = \frac{A_C}{|\Gamma|} \left(\frac{A_C}{r_0^2}\right)^{\frac{d-2}{2}}
\end{equation}
we eliminate this factor from the loop equation altogether. But keep in mind that reaching the equilibrium PDF we are investigating would take large time  $T_0$  at $d >2 $. ( we reinserted here missing dimensional factors ).

After dropping the factor $r_0$ in 3 dimensions:
\begin{equation}\label{LoopEq1}
    \oint_C d r_i \int_{S_C} d \sigma(r') \frac{ (r'_i - r_i)n_k(r)n_k(r') - n_i(r')(r'_k - r_k) n_k(r) }{|\vec r' - \vec r |^3}=0
\end{equation}
This is the final form of the Loop Equation for stationary PDF depending of the minimal area. 

\section{Exact Solution for Flat Loop}

Here is the biggest news of this paper: the minimal surface \textbf{exactly} solves the loop equation for a flat loop (and as a corollary, for whole 2D turbulence problem).

The thing is, that at flat loop (in the $x,y $ plane) the minimal surface is flat as well so that both $\vec n(r) = \vec n(r') = (0,0,1)$ which makes the second term in (\ref{LoopEq1}) zero. As for the first term, it reduces to the gradient and vanishes after integration over closed loop:
\begin{equation}
    \oint_C d r_i \int_{S_C} d \sigma(r')\frac{ (r'_i - r_i) }{|\vec r' - \vec r |^3} \propto  \int_{S_C} d \sigma(r')\oint_C d r_i \partial_{r_i}\frac{ 1}{|\vec r' - \vec r |} =0
\end{equation}

So, we claim that this is not just an asymptotic solution for the tails of the PDF, this is exact solution.

In case of 2D turbulence, where all loops are planar and vorticity is a pseudoscalar and the normal vector is $n =\pm 1$ depending on the orientation of the loop, the loop equation for the minimal surface Anzatz reads
\begin{equation}
    \oint_C d r_i \int_{S_C} d \sigma(r')\frac{ (r'_i - r_i) }{|\vec r' - \vec r |^2} \propto  \int_{S_C} d \sigma(r')\oint_C d r_i \partial_{r_i}\ln |\vec r' - \vec r | =0
\end{equation}

Let us come back to exact solution of the loop equation  in three dimension for a planar loop.
In that case it must apply to the moments of the circulation, in particular, to the second moment
\begin{equation}
    \left<\Gamma^2\right> = A_C \int_{-\infty}^{\infty} d \gamma \gamma^2 \Pi(\gamma)
\end{equation}
This formula raised objections\footnote{Sasha Polyakov, private communication.}: you can explicitly compute this double integral
\begin{equation}\label{Moment2VV}
    \left<\Gamma^2\right> = \oint_C d r_i \oint_C d r'_j \left< v_i(r) v_j(r')\right>
\end{equation}
taking the scaling law $\left< v_i(r) v_j(r')\right> \propto \delta_{i j} |r-r'|^{2 \alpha -1}$ where $\alpha$ is the scaling index of circulation in terms of the area. Integral looks nothing like an area and, say, for the rectangle it shows manifest dependence upon the aspect ratio at fixed area.

Our answer is very simple: this is so for Kolmogorov index $\alpha = \frac{2}{3}$ as well as any other index except our prediction\footnote{Note that the argument of \cite{M19b} applies to arbitrary dimension of space, as long as the vorticity surface was 2-dimensional.} $\alpha = \frac{1}{2}$. In this exceptional case the second moment can be directly proven to be equal to the area.

Namely, in case $2\alpha = 1$ the velocity has zero dimension, so its correlator is proportional to $\ln |\vec r-\vec r'|$.
The double loop integral by Stokes theorem reduces to double area integral
\begin{align}
    &\left<\Gamma^2\right> = \oint_C d r_i \oint_C d r'_j \left< v_i(r) v_j(r')\right>=\\
    &\int_{S_C} d \sigma(r) \int_{S_C} d \sigma(r')\left< \omega_3(r) \omega_3(r')\right> \propto \\
    &\int_{S_C} d \sigma(r) \int_{S_C} d \sigma(r') \nabla^2 \ln |\vec r-\vec r'| \propto \\
    &\int_{S_C} d \sigma(r) \int_{S_C} d \sigma(r') \delta(\vec r-\vec r')  = A_C
\end{align}
Q.E.D.

We also claim that for our solution all higher moments are  proportional to powers of the area for the flat loop, though this is hard to verify directly, as we do not know the exact form of higher velocity correlations.

Note however, that the same , logarithmic velocity correlator in 3D space, as required to compute the second moment for non-planar loop, will no longer produce $\delta$ functions for vorticity correlations. We may use the Stokes theorem and integrate over some curved surface but vorticity correlator
\begin{equation}
    \left< \omega_i(r) \omega_j(r')\right> \propto \left(\delta_{i j} \partial^2 - \partial_i \partial_j \right)   \ln |\vec r - \vec r'|  
\end{equation}
will have long term tails $\propto \frac{1}{(\vec r - \vec r')^2} $ unless taken on a flat surface.
Therefore, our solution does not imply short range correlation of vorticity, just that it has scaling dimension $-1$.

Interesting property of our solution is that it leaves the scaling function arbitrary, as long as it depends upon minimal area. The dependence of higher correlations of vorticity in a background of circulation is uniquely expressed in terms of basic PDF scaling function, but this function remains arbitrary at this point.

We shall accept the solution for the planar loop and try to generalize it for the non-planar one.

\section{ Loop Equation in Quadratic approximation}

Let us introduce a local quadratic approximation to a surface, as before (with $K_1 = K, K_2= -K$), and $n(r')$ given by (\ref{Normal2}), with $z$ direction being the local normal vector at the contour. The curvature $K$ refers to the integration point $r$. 

As we found the minimal surface in Appendix H in \cite{M93} the conformal coordinates at the boundary are consistent with these $x,y$ in quadratic approximation: $x$ goes along the local tangent direction of the loop and $y$ goes inside the surface. As for $z$ it obviously goes along the normal to the surface $n(r)$. We take $r$ as an origin. The loop $C $ expands as a piece of parabola in $x y $ plane:
\begin{align}
    C_1 &= x \\
    C_2 &= b x^2\\
    C_3 &=0
\end{align}

Therefore the surface integral in quadratic approximation the integrand in $\oint_C d r_i $ becomes:
\begin{equation}
    \int_{-\infty}^{\infty} d y
    \int_{-\infty}^{\infty} x\,d x \,
    \theta\left(y - b x^2\right)
     \frac{
     1 + \frac{K^2}{2}\left(x^2-y^2\right)
     }
     {
     \sqrt{
     \left(x^2 + y^2\right)^3\left(1 + K^2 \left(x^2 + y^2\right)\right)
     }
     }
\end{equation}

This integral vanishes by reflection symmetry $x\rightarrow -x$.

However, there is no reason to expect this reflection symmetry to hold beyond quadratic approximation. Cubic terms in equation of the contour alone would destroy this reflection symmetry.

This leaves us in desperate need for a modification of an Area such that the integral in the loop equation will exactly cancel to zero for arbitrary loop.

\section{Minimal Area In Conformal Metric Field}
The computations in the previous Section  suggest the following "conformal" Anzatz
\begin{equation}\label{Conformal}
  A_C[\phi] =\min_{S_C} \int_{S_C} d \sigma_{i}(r_1) \int_{S_C} d \sigma_{j}(r_2) \delta_{i j} \Delta(r_1-r_2) \exp\left(\frac{1}{2}( \phi(r_1) + \phi(r_2))\right)
\end{equation}
where conformal metric $\phi(r)$ is some external field defined in all $R_3$, not just on the surface. 

The local limit of this functional $A_C[\phi]$ tends to the area in external conformal metric
\begin{align}
    &A_C[\phi] \underset{r_0\rightarrow 0}{\longrightarrow} \int_{S_C} d \sigma(r) \exp\left(\phi\left(R(x)\right)\right)\\\label{LocalArea}
   &d \sigma(r) = \sqrt{\left(d \sigma_i(r)\right)^2} = d^2 x \sqrt{\det G}
\end{align}
with induced metric tensor, corresponding to parametric equation $\vec r = \vec R(x), x = (u,v)$ of the surface $S_C$
\begin{equation}
    G_{a b}(x) = \partial_a R_\mu(x) \partial_b R_\mu(x)
\end{equation}
We derive equation for $\phi$ later  but now consider this Area in external field $A_C[\phi]$ as a functional of the surface at fixed external field $\phi(r)$.
First of all, one can verify that this is a positive definite functional, just as before
\begin{align}
    &\int_{S_C} d \sigma_{i}(r_1) \int_{S_C} d \sigma_{j}(r_2) \delta_{i j} \Delta(r_1-r_2)\exp\left(\frac{1}{2}( \phi(r_1) + \phi(r_2))\right)\\
    &\propto \int d^3k \exp\left(-\frac{k^2 r_0^2}{4\pi}\right) \left| \int_{S_C} d \sigma_{i}(r)\exp\left(i k r + \frac{1}{2}\phi(r)\right) \right|^2
\end{align}

Now, the Stokes condition will, as before, be satisfied in virtue of minimality. Still, it will be interesting to derive replacement of mean curvature  equation for the ordinary minimal surface.
Repeating above computations in local tangent plane we find here:
\begin{equation}\label{MeanCurv}
    K_1 + K_2 = n_i(x) \partial_i \phi(r)_{\vec r =\vec R(x)}
\end{equation}
On the left we have mean curvature at the surface, and on the right we have normal derivative of the conformal metric field projected at the surface. 

This means that we can keep the ordinary minimal surface, with $ K_1 + K_2 =0$ and make sure that the conformal metric varies only along the surface but not in the normal direction. In the local tangent plane  we are using this means that $\phi(x,y,z) = \phi(x,y,0)$, so that it changes only in a local tangent plane but not in the normal direction.

Remember, this is just the Stokes condition, not yet the loop equation, but the net result is that we keep the minimal surface in a sense of being the surface of minimal area in ordinary induced metric. However, the Extended Area $A_C[\phi]$ has an extra factor $\exp(\phi(r))$ which varies along the surface.

For the loop equation to hold, the following condition must be valid:
\begin{equation}\label{SelfCons}
    \oint_C d r_i \exp\left(\phi(r)\right)\int_{S_C} d \sigma(r')\exp\left(\phi(r')\right) \frac{ (r'_i - r_i)n_k(r)n_k(r') - n_i(r')(r'_k - r_k) n_k(r) }{|\vec r' - \vec r |^3}=0
\end{equation}
We shall call it the self-consistency relation for the metric field. At given surface $S_C$ this is an equation for the metric field on the surface.
Thus we get closed set of integro-differential equations for parametric equation of surface $r_i =R_i(u,v)$  and the external conformal metric $\phi(u,v)$ which is two dimensional vector field. So, we have 2-dimensional surface embedded in 4-dimensional space $x,y,z,\phi$.

Note also, that there would now extra terms in the loop equation of the structure
\begin{equation}
    \int d^3 \rho \frac{\rho_j}{|\vec \rho|^3}\Delta(\vec \rho)
    \exp\left(\frac{1}{2}\phi(\vec r+\vec \rho)+ \frac{1}{2}\phi(\vec r)\right)
\end{equation}
coming from the second area derivative of generalized area $A_{S_C}[\phi]$. In presence of conformal metric field these terms no longer vanish by space symmetry, but the leading term at $r_0 \rightarrow 0$ will be a gradient which vanishes after loop integration\footnote{note also that extra condition $n_k \partial_k n(r)=0$ was not used here, as $n_k(r) d r_k =0$.}
\begin{equation}
   \oint_C d r_i \int d^3 \rho \frac{\rho_i}{|\vec \rho|^3}\Delta(\vec \rho)\exp\left(\frac{1}{2}\phi(\vec r+\vec \rho) + \frac{1}{2}\phi(\vec r)\right) \underset{r0\rightarrow 0}{\propto}  \oint_C d r_i\partial_i \phi(\vec r)\exp \left(\phi(\vec r)\right) = 0
\end{equation}
The next terms with $\partial^3 \phi , \left(\partial^2 \phi\right) \partial \phi , \left(\partial \phi\right)^3$ will already have $r_0^2$ in front of them, so they will be negligible compared to the leading $O(1)$ term in the loop equation.

Let us integrate the first term in (\ref{SelfCons}) by parts
\begin{align}
    &\oint_C d r_i \exp\left(\phi(r)\right)\int_{S_C} d \sigma(r')\exp\left(\phi(r')\right) \frac{ n_i(r')(r'_k - r_k) n_k(r) }{|\vec r' - \vec r |^3}=\\
    &\oint_C d r_i \exp\left(\phi(r)\right)\int_{S_C} d \sigma(r')\exp\left(\phi(r')\right) n_k(r)n_k(r')\partial_{r_i}\frac{1}{|\vec r' - \vec r |}=\\
    &-\oint_C d r_i \exp\left(\phi(r)\right)\left(\partial_{i}n_k(r) + n_k(r)\partial_{i}\phi(r)\right)\int_{S_C} d \sigma(r')\exp\left(\phi(r')\right)\frac{ n_k(r')}{|\vec r' - \vec r |}
\end{align}
Moving term with $\partial_i \phi(r)$ to the left and all remaining terms to the right we get
\begin{align}\label{LoopEq}
    &\oint_C d r_i \exp\left(\phi(r)\right)\partial_{i}\phi(r)\int_{S_C} d \sigma(r')\exp\left(\phi(r')\right)\frac{ n_k(r)n_k(r')}{|\vec r' - \vec r |}=\\
    &-\oint_C d r_i \exp\left(\phi(r)\right)\int_{S_C} d \sigma(r')\exp\left(\phi(r')\right) \left(\frac{ \left(n_i(r')-n_i(r)\right)(r'_k - r_k) n_k(r) }{|\vec r' - \vec r |^3} + \frac{\partial_{i}n_k(r)  n_k(r')}{|\vec r' - \vec r |}\right)
\end{align}
We used the fact $ d r_i n_i(r) =0$ to subtract $n_i(r)$ from $n_i(r')$ and remove spurious singularity in the surface integral.
At this point exact solution seems out of question - even the mean curvature equation is a problem which can only tackled by numerical minimization.

\section{Vorticity Correlations and Equation for Scaling Index}

Let us now repeat computation of vorticity correlations. The area derivatives are modified  in a trivial way (because regularized Area is quadratic functional of $ d \sigma_i$ we get factor of 2):
\begin{align}
    \frac{\delta A_C[\phi]}{\delta \sigma_i(r)} &= 2 n_i(r) \exp \left( \phi(r)\right)\\
    \int_{S_C} d \sigma_i(r) \frac{\delta A_C[\phi]}{\delta \sigma_i(r)} &= 2\int_{S_C} d \sigma_i(r) n_i(r)\exp \left( \phi(r)\right)\\
    &= 2  A_C[\phi]
\end{align}
so, we get the same equations for vorticity correlations, including the self-consistency equation for $\alpha$:
\begin{equation}
    \alpha = \frac{1}{2}
\end{equation}
All the vorticity correlations will acquire extra factors of $\exp ( \phi)$:
\begin{equation}\label{MultiNormal}
    \left < \vec \omega_1\dots \vec\omega_k \delta\left(\Gamma - \oint_C \vec v d\vec r\right)\right >  = \left|A_C[\phi]\right|^{-\frac{1}{2}}\vec n_1 \exp ( \phi_1)\dots \vec n_k \exp ( \phi_k)A_C[\phi]^{-\frac{k}{2}}\Omega_k\left(\Gamma A_C[\phi]^{-\frac{1}{2}}\right)
\end{equation}

So, the formulas look the same and the scaling functions $\Omega_k\left(\Gamma A_C[\phi]^{-\frac{1}{2}}\right)$ satisfy the same recurrent equations and results expressing them in terms of integrals of $\Omega_0(\gamma) = \Pi(\gamma)$ stay the same as in \cite{M19b}.
The only thing which really changed is extra dependence of the coordinates provided by conformal metric factor $\exp(\phi)$.
So, until we solve these heavy equations for $\phi$ and $\vec R(u,v)$ we cannot predict spatial dependence of vorticity correlations.

\section{Circulation Moments and Orientation Reversal Symmetry}

So, it looks like our equations (\ref{MeanCurv}), (\ref{LoopEq})  provide both the Stokes condition and the loop equation without any assumptions about the PDF tails.

In that case they must be applicable to the moments of circulation
\begin{equation}\label{Moments}
    \left < \Gamma^k
    \right > = \int_{-\infty}^{\infty} d \Gamma \Gamma^k P(\Gamma,C) 
    =\left(A_C[\phi]\right)^{+\frac{k}{2}}\int_{-\infty}^{\infty} d \gamma \gamma^k\Pi(\gamma)
\end{equation}
Here is some general argument\footnote{Sasha Polyakov, private communication.} that the odd moments must vanish.
When the loop $C$ changes its orientation $\vec C(t) \rightarrow \vec C(-t)$ , the vorticity $ \Gamma$ changes sign but the minimal area apparently does not.

Well, with conformal metric present this is not so simple anymore. 
The odd moments contain multiple contour integrals of velocity, or equivalently, multiple surface integrals of vorticity:
\begin{equation}
    \left < \Gamma^k
    \right > = \int_{S_C} d \sigma_{i_1}(r_1) \dots \int_{S_C} d \sigma_{i_k}(r_k) \left< \omega_{i_1}(r_1) \dots \omega_{i_k}(r_k)\right>
\end{equation}
The orientation symmetry means here that with change of the contour orientation the same happens with orientable surface $S_C$ after which each surface element $d \sigma_{i}(r)$ changes sign due to reflection of the normal vector. The expectation values of products of vorticities clearly do not depend upon the surface in Stokes theorem so that the RHS  changes sign for odd $k$.

Now, in our solution (\ref{Moments}) the same thing would happen if we would multiply $A_C[\phi]$ by $e^{2\pi i}$. This would mean switching to the negative branch of the square root. Such a multiplication can be achieved by shifting conformal metric $\phi(r) \rightarrow \phi(r) + 2\pi i $, which is of course an equivalent conformal metric, equally satisfying the loop equation. In other words, changing orientation of minimal surface must be accompanied by identity phase shift of conformal metric.

Note that this is an independent reason why the scaling index must be half-integer, otherwise the identity shift $\phi \rightarrow \phi + 2 \pi i$ would produce unacceptable complex phase factor $ \exp\left(2 i\pi \alpha\right)$.

So, one can object that orientation  reflection either do not apply to our asymptotic solution at $ |\Gamma| \gg \nu$ or must be accompanied by changing the sign of $\sqrt{A_C}$.

Here is, however a much more direct argument in favor of vanishing odd  moments.
Taking the formulas for multiple vorticity correlations in presence of circulation $\Gamma$ in a limit $\Gamma\rightarrow 0$ in \cite{M19b} we find that 
\begin{align}
    \Omega_{2k}(0^+) &= B_k \int_0^{\infty} \gamma^{2k-1} \Pi(\gamma) \, d\gamma\\
    \Omega_{2k}(0^-) &= B_k \int_0^{-\infty} \gamma^{2k-1} \Pi(\gamma) \, d\gamma
\end{align}
with some numbers\footnote{These numbers can be computed from the recurrent equations \cite{M19b}  for $\Omega_k$ by choosing $\Pi(\gamma) = \frac{1}{2}\exp(-|\gamma|)$.}$B_k \neq 0$.
Subtracting these two equations we find
\begin{equation}
    \Omega_{2k}(0^+)- \Omega_{2k}(0^-) = B_k \int_{-\infty}^{\infty} \gamma^{2k-1} \Pi(\gamma) \, d\gamma\\
\end{equation}

So, we conclude that either the limits $\Omega_{2k}\left(0^{\pm}\right)$ are different or the odd moments vanish. Can we prove that there is no discontinuity at $\Gamma\rightarrow 0^\pm$ in the vorticity correlations? 

Remember, we are talking about the limit of $\nu \ll \Gamma \ll \sqrt{A_C} $ on the positive side and the limit $\nu \ll -\Gamma \ll \sqrt{A_C} $ on the left side. This is not a discontinuity in mathematical sense, because there is a region $\Gamma \sim \nu$ in the way from positive to negative $\Gamma$. 

Rather than discontinuity at $\Gamma =0$ these are two different asymptotic regimes in a theory where the symmetry between positive and negative $\Gamma$ is known to be broken by the energy cascade followed by dissipation as viscous scales. Mathematically, this is Stokes phenomenon for our asymptotic solution at $|\Gamma| \gg \nu$.

We can imagine the regime change somewhere between $ \Gamma \sim \nu $ and some larger scale, where the PDF changes its scale from K41 $\Gamma \sim \left(\mathcal{E} A_C^2\right)^{\frac{1}{3}}$ to our regime $\Gamma \sim \sqrt{A_C}$. The PDF inside K41 region is asymmetric, and in particular, $\left<\Gamma^3\right>_{K41} \propto \mathcal{E} A_C^2 \ne 0$. 

This value contributes to total integral of the third moment. At large $A_C$ this contribution grows faster than our $A_C^{\frac{3}{2}}$ so that either our regime is before K41 or else there is a contribution from the transient region between K41 and our regime, which cancels that large K41 contribution at very large $A_C$. It could cancel it completely and then we will have vanishing odd moments or it can cancel K41 down to $A_C^{\frac{3}{2}}$ in which case there will be an asymmetry left.

 So, we leave this question open.
 
Our solution is essentially different from Kolmogorov cascade. Presumably, it holds at large enough scales in isotropic turbulence, and it is very different from Gaussian -- with almost linear exponential tails according to \cite{S19}. 
These experiments  seem to confirm area law, though not at all scales and all circulations.

The second moment $\left<\Gamma^2\right>$ in these experiments scales as K41 with area, which means it must behave as the integral (\ref{Moment2VV}) with $\alpha= \frac{2}{3}$. I computed this integral in \Mathematica\, for rectangle as a function of its perimeter for unit area. When perimeter goes up from $4$ to $6$ which is the range of variations of rectangular loop in \cite{S19} the ratio of second moment for rectangle to the same for a unit square goes down from $1$ to $0.8983$. The variations observed in these experiments\footnote{Kartik P. Iyer, private communication.} are consistent with these explicit computations .

So, observed weak dependence of the second moment of the aspect ratio of unit area rectangle does not prove the area law, it rather only confirms the theoretical dependence of this moment of the aspect ratio for a K41 index.
However, the Area law can also be compared with experiments at larger moments, which corresponds to larger $\Gamma$, deeper in its inertial interval $|\Gamma| \gg \nu$. The dimensionless differences
\begin{equation}
    \frac{\ln \left<|\Gamma|^p\right>}{p} -  \frac{\ln \left<|\Gamma|^{p-1}\right>}{p-1}
\end{equation}
must be checked for their perimeter dependence.
According to experiments, asymptotically, at large  $A_C$ the moments $\left< \Gamma^p\right>$ grow  as $\sqrt{A_C}^{\lambda(p)}$ with $\lambda(p) \approx p + 0.92 \ln p$.
The logarithmic correction to linear growth must be attributed to the transition from Kolmogorov cascade to vorticity surface.

This transition remains a challenge for the future.

\section{Conclusion}
The main result of this work is exact solution of the loop equation for arbitrary flat loop, which includes 2D turbulence as a whole. 
With scaling index $\alpha = \frac{1}{2}$ as derived in the previous work, the minimal area solves the loop equations and passes the known tests for the circulation moments.
For the general, non-planar loop, the equation for the minimal surface stays the same (mean curvature equals zero), and we derive integral equation  (\ref{LoopEq}) for the self consistent conformal metric.
The correlation scaling functions also stay the same, and local vorticity in expectation values is still directed along the normal to geometric minimal surface. The coordinate dependence of vorticity correlations gets modified by extra factors of conformal metric $\exp(\phi)$. In other words, we have a 2d-surface in 4 dimensional space $x,y,z,\phi$ with dependence of the $x,y,z$ coordinates given by the ordinary minimal surface, and the last coordinate changing along this minimal surface in case this surface is curved.

 \section{Acknowledgements}
 I am indebted to Sasha Polyakov and Victor Yakhot for stimulating discussions. 
 
\bibliographystyle{unsrt}  


\end{document}